\documentclass[aps,groupedaddress,twocolumn]{revtex4-1}

\usepackage{amsmath}
\usepackage{amsfonts}
\usepackage{amssymb}
\usepackage{graphicx}
\usepackage{verbatim}
\usepackage{xcolor}
\usepackage[normalem]{ulem}
\usepackage{appendix}
\newcommand{\ts}{\textsuperscript}

\usepackage[english]{babel}
\makeatletter
\def\bbl@set@language#1{%
  \edef\languagename{%
    \ifnum\escapechar=\expandafter`\string#1\@empty
    \else\string#1\@empty\fi}%
  \@ifundefined{babel@language@alias@\languagename}{}{%
    \edef\languagename{\@nameuse{babel@language@alias@\languagename}}%
  }%
  \select@language{\languagename}%
  \expandafter\ifx\csname date\languagename\endcsname\relax\else
    \if@filesw
      \protected@write\@auxout{}{\string\select@language{\languagename}}%
      \bbl@for\bbl@tempa\BabelContentsFiles{%
        \addtocontents{\bbl@tempa}{\xstring\select@language{\languagename}}}%
      \bbl@usehooks{write}{}%
    \fi
  \fi}
\newcommand{\DeclareLanguageAlias}[2]{%
  \global\@namedef{babel@language@alias@#1}{#2}%
}
\makeatother
\DeclareLanguageAlias{en}{english}
\begin{document}
\title{Wannier quasi-classical approach to high harmonic generation in semiconductors}

\author{A. M. Parks\ts{1}}
\email[]{andrew.parks@uottawa.ca}
\author{G. Ernotte\ts{1}}
\email[]{gerno013@uottawa.ca}
\author{A. Thorpe\ts{1}}
\author{C. R. McDonald\ts{1}}
\author{P. B. Corkum\ts{1}}
\author{M. Taucer\ts{1}}
\author{T. Brabec\ts{1}}
\affiliation{\ts{1}Department of Physics, University of Ottawa, Ottawa, ON K1N 6N5, Canada}
% \affiliation{\ts{2}National Research Council of Canada, Ottawa, Ontario, Canada K1A 0R6}

\date{\today}

\begin{abstract}
\noindent
We develop a quasi-classical theory of high harmonic generation in semiconductors based on an interband current that has been transformed from Bloch to Wannier basis. 
The Wannier quasi-classical approach reveals a complete picture of the mechanisms shaping high harmonic generation, such that quantitative agreement with full quantum calculations 
is obtained. The intuitive picture revealed by quasi-classical wavepacket propagation will be helpful in the interpretation and design of high harmonic and attosecond experiments. 
Beyond that, the capacity to quantitatively model quantum dynamics with classical trajectories should prove useful for a wider spectrum of condensed matter research, including coherent 
control, transport theory, and strong field physics. 
\end{abstract}

\maketitle

\section{Introduction}

\noindent
High harmonic generation (HHG) in solids has been demonstrated in a wide range of materials \cite{Ghimire2011, Zaks2012, Schubert2014, Hohenleutner2015, Luu2015, Garg2016, Vampa2015_2, 
Vampa2015_3, Liu2016, You2017, Banks2017, Worner2018, Uzan2019}; it has caught attention as a source for ultrashort xuv radiation and as a tool to measure ultrafast dynamics 
and structural properties, such band structure \cite{Luu2015, Vampa2015_2} and the Berry curvature \cite{Banks2017, Worner2018}. HHG in solids is driven by interband and 
intraband currents. While the interband current is more dominant in wide-band materials, such as semiconductors \cite{Vampa2015_2}, HHG in narrow-band dielectrics is driven more by the 
intraband current \cite{Luu2015}. This work focuses on interband HHG in wide-band materials. 

Although some experimental features can be reasonably well reproduced by numerical models \cite{Tancogne2017, Wu2017, Jiang2018, Li2019}, a thorough understanding of all the components 
shaping harmonic spectra is still missing. This inhibits progress in optimizing HHG as a radiation source and in further developing HHG as a diagnostic tool. 

The principal mechanism of interband HHG has been clarified by saddle point integration of the interband current derived in the Bloch basis \cite{Vampa2014}. Electron and hole are born at the
same lattice site in real space by tunnel ionization and quiver in the laser field. When they recollide at some lattice site, a harmonic photon is emitted. Its energy is equal to the bandgap 
at the crystal momentum of the electron-hole pair at recollision. Despite its merits, the Bloch quasi-classical model falls short of accounting for the lattice structure; quantum mechanics
allows recombination of electrons and holes at different lattice sites, as was clearly demonstrated in recent work \cite{You2017, Osika2017, Yue2020}. 

Here we develop a generalized quasi-classical approach that accounts for the lattice structure; this is achieved by transforming the interband current from Bloch to Wannier basis followed by 
saddle point integration. The basis change has a substantial effect. The resulting Wannier quasi-classical (WQC) model is found to be in quantitative agreement with quantum 
calculations. So far, quasi-classical $k$-space analysis has been used to qualitatively investigate strong field effects in gases and in the condensed matter phase; quantitative 
agreement has not been demonstrated yet. Whether quantitative agreement can be obtained in the Bloch basis remains to be seen, however the richer physics revealed by the WQC picture indicates 
that this might not be the case. The more refined WQC picture arises from the fact that the transition dipole moment enters the classical action in the exponent, and therewith the saddle point 
equations. 

The quantitative agreement with full quantum calculations suggests that the physical picture for HHG in semiconductors revealed by the WQC analysis is complete. An electron and 
hole can ionize and recombine at different lattice sites with a probability determined by the tunneling exponent and Wannier dipole moments; birth and recombination sites are connected by
classical trajectories; quantum effects are included by a quadratic expansion of the classical action about the classical trajectories. Beyond that, our WQC analysis allows unprecedented 
insight into the real-space aspects of tunnel ionization in solids; it gives access to the tunnel ionized wavefunction in real space and therewith, to the birth location of the
electron hole pair. 
 
More generally, our analysis opens an avenue for modeling quantum dynamics of wavepackets by propagating classical trajectories. This is potentially relevant for a wide spectrum of 
applications ranging from strong field physics to transport phenomena \cite{Datta1997,Ludwig2019} and coherent control \cite{Stevens2005, Muniz2014}. On a fundamental level, the WQC approach 
could open an alternative pathway to modeling noise and few electron-hole dynamics in solids; as propagation from initial to final Wannier wavepacket is done by classical trajectories, the 
space in between does not need to be resolved in contrast to a full quantum approach. 

\section{Theory}

\subsection{Two Band WQC Model}

Our formalism is developed for a 3D, two-band model. We first summarize derivation of HHG in the Bloch basis \cite{Vampa2014}; it starts from the time-dependent Hamiltonian $H(t) = H_{0} + 
{\bf x}\cdot{\bf F}(t)$; ${\bf F}(t)$ represents the laser field; $H_{0}$ is the unperturbed lattice Hamiltonian with Bloch eigenstates $\Phi_{{m},{\bf k}}({\bf x}) = 1/\sqrt{V} 
u_{m,{\bf k}}({\bf x}) \exp(i{\bf k\cdot x})$ and with energies $E_{m}({\bf k})$ in band $m$ with crystal momentum $\bf k$; the band index $m=v,c$ refers to valence and conduction band,
respectively; $u_{{m},{\bf k}}$ is the periodic part of the Bloch function, $\langle \Phi_{{m},{\bf k}} \vert \Phi_{{m},{\bf k}} \rangle = 1$, and $\langle u_{{m},{\bf k}} 
\vert u_{{m},{\bf k}} \rangle = \upsilon$. Finally, $V= N \upsilon$ is the volume of the solid, with $N$ and $\upsilon$ the number and volume of primitive unit cells. Hartree atomic 
units are used, unless otherwise noted. 

In the presence of the laser field the wavefunction becomes time-dependent. In the length gauge it is represented as
\begin{equation}
\Psi({\bf x},t)  = \sum_{m=v,c} \int_{\rm BZ} a_{m}({\bf k}, t) \Phi_{{m},{\bf k}}({\bf x}) \, d^3{\bf k} \text{,}
\label{ansatz}
\end{equation}
where $a_{m}({\bf k},t)$ are the probability amplitudes and integration is over the full Brillouin zone (BZ). As initial conditions we choose an empty conduction band $a_{c}({\bf k},t=0) 
= 0$, and a filled valence band, $a_{v}({\bf k},t=0) = 1/\sqrt{V_{BZ}}$, where $V_{BZ}$ is the Brillouin zone volume. The Ansatz (\ref{ansatz}) is substituted into the time-dependent 
Schr\"odinger equation, and the interband polarization and current are found to be \cite{Vampa2014}
\begin{subequations}
\label{jerom}
\begin{align}
\mathbf{p}_{er}(t) & \!=\! -i\!\!\int_{\rm BZ} \!\!\!\!\! d\mathbf{k} \, \mathbf{d}(\mathbf{k}) \!\!\! \int_{-\infty}^{t} \!\!\!\!\!\!\! dt' \mathbf{F}(t')\! \cdot\!\mathbf{d}^*[\mathbf{k}(t',t)] 
e^{-i S(\mathbf{k},t',t)}\!+\!\text{c.c.} \label{pert} \\
\tilde{\mathbf{j}}_{er}(\omega) & = \! i\omega \!\! \int_{-\infty}^{\infty} \!\!\! dt e^{-i\omega t}\mathbf{p}_{er}(t)   \label{jerom1} 
\end{align}
\end{subequations}
with $S(\mathbf{k},t',t) = \int_{t'}^{t} \varepsilon(\mathbf{k}(t'',t) ) dt'' - i (t-t')/T_2$, $T_2$ the dephasing time, $\mathbf{k}(t',t) = \mathbf{k} + \mathbf{A}(t) - 
\mathbf{A}(t')$ with $\mathbf{A}(t)$ the vector potential satisfying $\mathbf {F}=-\partial_t\mathbf{A}$, and $\varepsilon = E_c-E_{v}$.  Here, we have used the relation \cite{Blount1962} $\langle\Phi_{m,\mathbf{k}}|\mathbf{x}|\Phi_{m',\mathbf{k}'}\rangle=\delta(\mathbf{k}-\mathbf{k}')[i\delta_{m,m'}\nabla_{\mathbf{k}}+\mathbf{d}_{mm'}(\mathbf{k})]$, with $\mathbf{d}_{mm'}(\mathbf{k})=i\langle u_{m,\mathbf{k}}|\nabla_\mathbf{k}|u_{m,\mathbf{k}}\rangle$ the transition dipole moment.  For a two-band system, we denote
\begin{align}
\mathbf{d}(\mathbf{k})= \mathbf{d}_{vc}(\mathbf{k})= i\langle u_{v,\mathbf{k}}|\nabla_\mathbf{k}|u_{c,\mathbf{k}}\rangle \text{,}
\label{kdipole}
\end{align}
and we assume a centro-symmetric system for which the diagonal elements $\mathbf{d}_{mm}(\mathbf{k})$ can be set to zero \cite{Li2019_phaseinv}.

In the following we will translate HHG, as described by the interband current of (\ref{jerom}), from $k$-space to real space by using Wannier functions. 
The Bloch and Wannier basis functions are connected by a Fourier transform according to \cite{Haug&Koch}
\begin{subequations}
\label{ft}
\begin{align}
u_{m,\mathbf{k}}(\mathbf{x}) & = \sum_{j} w_{m}(\mathbf{x} - \mathbf{x}_j) e^{-i\mathbf{k}\cdot(\mathbf{x}-\mathbf{x}_j)}  \label{wtob} \\ 
w_{m}(\mathbf{x} - \mathbf{x}_j) & = \frac{1}{\upsilon} \int_{\rm BZ} u_{m,\mathbf{k}}(\mathbf{x}) e^{i\mathbf{k}\cdot(\mathbf{x}-\mathbf{x}_j)} 
d\mathbf{k} \text{.} \label{btow}
\end{align}
\end{subequations}
Here, $w_{m}(\mathbf{x} - \mathbf{x}_j)$ is the Wannier function of band $m$ corresponding to the primitive unit cell at position $\mathbf{x}_j$.
By virtue of (\ref{btow}), the initial wavefunction, 
\begin{align}
\Psi({\bf x},0)  = \int_{\rm BZ}d{\bf k} \Phi_{{v},{\bf k}}({\bf x}) a_{v}({\bf k},t=0) \,  = w_{m}(\mathbf{x}) \text{,}
\label{initial}
\end{align}
corresponds to the Wannier function at position $\mathbf{x}_j = 0$. HHG can start from any other site $\mathbf{x}_j$. The initial Wannier function can be shifted to $\mathbf{x}_j$ by 
setting $a_{v}({\bf k},t=0) = \exp(-i \mathbf{k}\cdot \mathbf{x}_j)$. As all lattice sites are identical, it is sufficient to investigate $\mathbf{x}_j = 0$. 

In order to translate the interband current (\ref{jerom}) into real space, the Bloch functions in the transition dipole moment (\ref{kdipole}) are replaced by the Wannier functions 
with the help of relation (\ref{wtob}). This leads to 
\begin{align} 
& \mathbf{d}(\mathbf{k}) \! = \sum_{j,k} \int_{\upsilon} \!\! w_{v}^*(\mathbf{x}-\mathbf{x}_k) [\mathbf{x}-\mathbf{x}_j] w_{c}(\mathbf{x}-\mathbf{x}_j) 
e^{i\mathbf{k}\cdot(\mathbf{x}_j-\mathbf{x}_k)} d\mathbf{x} \nonumber \\
~ & = \sum_{j,l} \int_{\upsilon} \!\! w_{v}^*(\mathbf{x}-(\mathbf{x}_j+\mathbf{x}_l)) [\mathbf{x}-\mathbf{x}_j] w_{c}(\mathbf{x}-\mathbf{x}_j) 
e^{-i\mathbf{k}\cdot\mathbf{x}_l} d\mathbf{x} \nonumber \\
~ & = \sum_{l} e^{-i\mathbf{k}\cdot\mathbf{x}_l} \! \int_{V} \!\! w_{v}^*(\mathbf{x} - \mathbf{x}_l) \, \mathbf{x} \, w_{c}(\mathbf{x}) d\mathbf{x} 
= \sum_{l} \mathbf{d}_l e^{-i\mathbf{k}\cdot\mathbf{x}_l} \text{,}
\label{dipow}
\end{align}
where the second line was obtained by setting $\mathbf{x}_k = \mathbf{x}_j + \mathbf{x}_l$ and by replacing summation index $k$ with $l$ in the first line. 
Also, note that performing $\sum_j$ in the second line changes the integration volume from a unit cell to the whole crystal volume. The Wannier dipole moments are equivalent 
to the Fourier series expansion coefficients of the Bloch dipole moment $\mathbf{d}(\mathbf{k})$. Interpreted in real space, the Wannier dipole moment $\mathbf{d}_l$ describes a transition 
where an electron is born $l$ lattice cells away from the hole. Bloch and Wannier dipole moments are not unique; $\Phi_{{m},{\bf k}} \rightarrow \Phi_{{m},{\bf k}} 
\exp[i \alpha(\mathbf{k}) ]$ is also an eigenfunction for any real function $\alpha$ that is periodic in $k$-space. Although the full equations, including the diagonal
dipole elements $\mathbf{d}_{mm}$, are gauge invariant \cite{Blount1962,Li2019_phaseinv}, it is computationally advantageous to choose strongly confined Wannier basis functions 
\cite{Kohn1959, Mostofi2014} in order to keep the number of relevant lattice sites small. In the 1D examples discussed further down we chose maximally localized Wannier basis functions \cite{Kohn1959} for which $\mathbf{d}_{mm} = 0$.

Inserting (\ref{dipow}) into (\ref{jerom}), the interband current follows as 
\begin{subequations}
\label{jeromw}
\begin{align}
\tilde{\mathbf{j}}_{er}(\omega) & =\! \sum_{j,l} \left\{ \mathbf{d}_j [\mathbf{d}_l^*\cdot \mathbf{T}_{jl}(\omega)]\! -\! \mathbf{d}_j^\ast[\mathbf{d}_l\cdot 
\mathbf{T}_{jl}^\ast(-\omega)]\right\} \nonumber \\ 
&=\! \sum_{j,l}\left[ \mathbf{P}_{jl}(\omega) - \mathbf{P}^\ast_{jl}(-\omega)\right] \text{,} \label{jeromw1} \\ 
\mathbf{T}_{jl}(\omega) & = \omega\!\! \int_{\rm BZ} \!\!\!\! d\mathbf{k} \!\! \int_{-\infty}^{\infty} \!\!\!\!\! dt \! \int_{-\infty}^{t} 
\!\!\!\!\!\! dt' \mathbf{F}(t') e^{i \varphi(\mathbf{k},t',t,\mathbf{x}_l,\mathbf{x}_j) } \text{,} \label{jeromw2} 
\end{align}
\end{subequations}
Here $\varphi = -S(\mathbf{k},t',t) -\omega t + \mathbf{k}\cdot (\mathbf{x}_l - \mathbf{x}_j) + [\mathbf{A}(t)-\mathbf{A}(t')]\cdot \mathbf{x}_l$; $\mathbf{P}_{jl}(\omega)$ represents the probability 
amplitude that the harmonic $\omega$ is generated by an electron-hole pair that is born with a relative distance $|\mathbf{x}_l|$ between electron and hole and later recombines with relative distance $|\mathbf{x}_j|$, and the propagator $\mathbf{T}_{jl}$ describes the evolution between 
$\mathbf{d}_l^*$ and $\mathbf{d}_j$. 

\subsection{Saddle Point Integration}

\begin{figure*}[t]
\includegraphics[width=16cm]{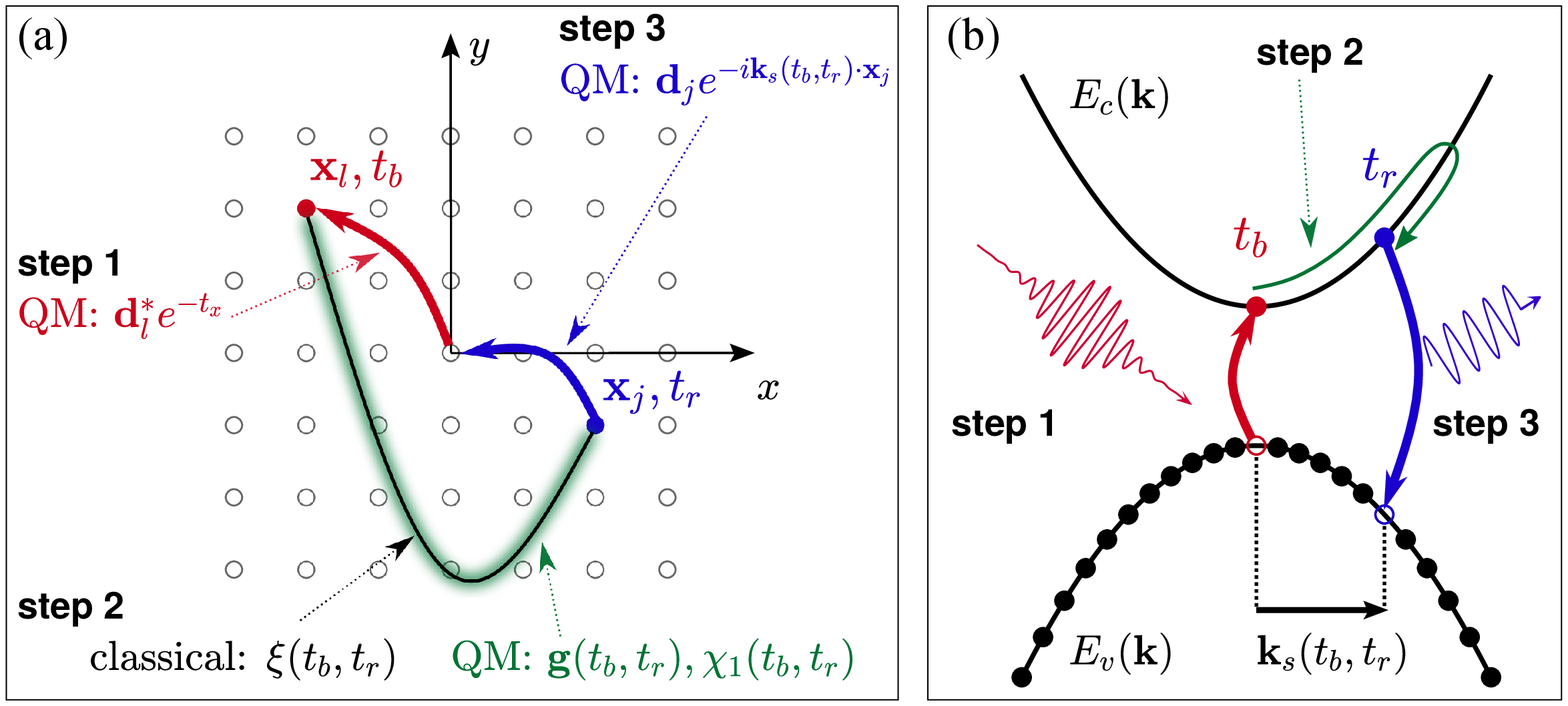} 
\caption{\label{fig1} Schematic of the WQC picture of interband HHG. 
(a) real space picture for a model 2D lattice: empty circles denote centers at atomic unit cells at which Wannier basis functions are located. Distances shown refer to relative distance 
between electron hole pair. Classical and quantum mechanical (QM) processes are indicated. The dotted arrows point to the probability amplitudes of the individual processes. In addition 
the phase $\chi_1$ picked up along the classical trajectory is indicated. HHG takes place in 3 steps. 
1) An electron initially at the valence Wannier site $\mathbf{x}_0$ is born at $\mathbf{x}_0+\mathbf{x}_l$ creating an electron-hole pair $l$ lattice sites apart (red arrow) with 
ionization amplitude $\propto \mathbf{d}^\ast_l\exp[-t_x]$. 
2) Then it propagates in the laser field along the classical trajectory $\boldsymbol{\xi}(t_b,t_r)$; the green shaded area indicates the quasi-classical contribution $\mathbf{g}(t_b,t_r)$
that comes from the Gaussian expansion of the propagator about the classical trajectory $\boldsymbol{\xi}$. 
3) Electron and hole revisit each other and recombine $j$ lattice sites apart with $\mathbf{d}_j$ the recombination dipole (blue arrow). 
(b) $\mathbf{k}$-space picture: full and empty circles in valence band indicate filled states and empty states (holes), respectively;
1) electron-hole pairs are born at the $\Gamma$ point ($\mathbf{k}=0$); 
2) the laser field drives them in reciprocal space (green arrow); 
3) they recombine at some different $\mathbf{k}_s$. }
\end{figure*}

\noindent 
The integrals in (\ref{jeromw2}) are solved by saddle point integration. The saddle point equations, 
\begin{subequations}
\label{saddle}
\begin{align}
&  \varepsilon[ \mathbf{k}(t',t) ] + \mathbf{F}(t')\cdot \mathbf{x}_l = 0 \label{stp} \text{,} \\
&  \varepsilon(\mathbf{k}) - \mathbf{F}(t) \cdot[ \boldsymbol{\xi}(t',t) - \mathbf{x}_l ] = \varepsilon(\mathbf{k}) + \mathbf{F}(t)\cdot \mathbf{x}_j = \mp \omega \label{st} \text{,} \\
&  \boldsymbol{\xi}(t',t) = \mathbf{x}_l -  \mathbf{x}_j  \label{sk} \text{,}
\end{align}
\end{subequations}
result from $\partial \varphi / \partial \mu\!\!=\!\!0$ with $\mu = t', t, \mathbf{k}$, respectively. The field quiver motion between times $t'$ and $t$ is given by the distance 
$ \boldsymbol{\xi}(\mathbf{k},t',t) = \int_{t'}^t \mathbf{v}( \mathbf{k}(t'',t) ) \, dt''$, where $\mathbf{v}(\mathbf{k}) = \boldsymbol{\nabla}_{\mathbf{k}} \varepsilon$ is the band 
velocity. Note that the classical action depends on the difference between conduction and valence band. As a result, the above quantities represent the difference between electron and 
hole band velocity and excursion distance. Finally, the $\mp$ in (\ref{st}) accounts for the complex conjugate term in (\ref{jerom1}). 

The set of equations (\ref{saddle}) are solved for a linearly polarized laser field $\mathbf{F} = F \hat{\mathbf{x}}$; further $\mathbf{A} = A \hat{\mathbf{x}}$ and $k_x = k$. 
The solutions of the saddle point equations are denoted by $t' = t_b + i \delta$, $t = t_r$, $\mathbf{k}_{s}$. For $\delta \ll 1$, (\ref{stp}) can be solved analytically; it 
determines the saddle point momentum $\mathbf{k}_s = (k_{s},k_{ys},k_{zs}) = (A(t_b) - A(t_r),0,0)$, as well as 
\begin{align}
\delta = \sqrt{ \frac{2 (E_g + F(t_b) x_l)}{\beta_{xx}(0) F^2(t_b)} } \text{,}
\label{delta}
\end{align}
where we have approximated the bandgap as
\begin{align}
\label{Equad}
\varepsilon(\mathbf{k})\approx E_g + \frac{1}{2}\sum_{i,j} k_i k_j\beta_{ij}(0)
\end{align} with $i,j = x,y,z$;  $\beta_{ij}(\mathbf{k}) = \partial^2 \varepsilon / \partial k_i \partial k_j$ the inverse mass tensor; and $E_g$ the minimum bandgap. The positive sign in (\ref{delta}) is chosen to obtain an 
exponentially decaying tunneling rate. 

The two remaining saddle point equations (\ref{st}) and (\ref{sk}) determine $t_b$ and $t_r$. They have to be solved numerically for each possible birth site $\mathbf{x}_l$ and recombination site $\mathbf{x}_j$; for instance, by running through $t_b$ and finding all $t_r(t_b)$'s that fulfill 
(\ref{sk}). From those, the pairs $[t_b, t_r](\omega)$ are selected that produce a given harmonic $\omega$ via (\ref{st}). 
The physical implications of the saddle point equations are discussed at the end of this subsection. 

Next, the integrand of (\ref{jeromw2}) is evaluated at the saddle point, where the small imaginary birth time determines the tunneling exponent. Further, the phase $\varphi$ is expanded 
to second order, which gives the multivariate Gaussian integral 
\begin{align}
\int_{-\infty}^{\infty} d\mathbf{q} \exp((i/2)\mathbf{q}^T \mathcal{H} \mathbf{q}) = (2\pi)^{5/2} / \sqrt{-i \lvert \mathcal{H} \rvert} \text{,} 
\label{saddle2}
\end{align}
where $\mathbf{q} = (t',t,\mathbf{k})$, and $\mathcal{H}$ is the Hessian $\mathcal{H}_{ij} = \partial^2 \varphi / \partial_i \partial_j$ with $i,j \in \mathbf{q}$. The full expression for 
the determinant of the Hessian is provided in appendix \ref{app_Hess}. Putting everything together, we obtain the WQC propagator
\begin{subequations}
\label{WQCprop}
\begin{align}
\mathbf{T}_{jl} & = \!\!\!\!\!\! \sum_{[t_b,t_r](\omega,\mathbf{x}_l,\mathbf{x}_j) } \!\!\!\! \mathbf{g}(t_b+i\delta, t_r) \, e^{-t_x}  e^{-i \chi(t_b,t_r) + i \pi/4}  \text{,}
\label{tprop} \\
t_x & = \textrm{Im}[\varphi(t_b+i\delta)] \approx \frac{ \sqrt{2} [E_g + F(t_b) x_l]^{3/2}}{ [\beta_{xx}(0) F^2(t_b)]^{1/2} } \text{,} \label{tunnel} \\
\chi & = \int_{t_b}^{t_r} \varepsilon(A(t_b) - A(\tau)) d\tau + \omega t_r + \mathbf{k}_s\cdot \mathbf{x}_j \text{,} \label{phchi} 
\end{align}
\end{subequations}
where $\mathbf{g} = \omega \mathbf{F}(t_b+i\delta) (2 \pi)^{5/2} / \sqrt{\lvert \mathcal{H} \rvert} $ and to leading order the determinant from the Gaussian integral $\lvert \mathcal{H} 
\rvert \approx v_x(\mathbf{k}_s) f(t_b+i\delta,t,\mathbf{k}_s)$ \cite{Uzan2019}, see appendix \ref{app_Hess}. Further, it is convenient to split the phase in (\ref{phchi}) into $\chi = 
\chi_1 + \chi_2$, where $\chi_1 = \int_{t_b}^{t_r} \varepsilon[A(t_b) - A(\tau)] \, d\tau \,+\, \omega t_r$ contains the classical action and the harmonic frequency Fourier term. The second 
term is the Fourier term of the recombination dipole moment, $\chi_2 = \mathbf{k}_s \cdot\mathbf{x}_j$. 
The total probability amplitude
\begin{align}
\mathbf{P}_{jl}= e^{i\pi/4}\sum_{[t_b,t_r](\omega,\mathbf{x}_l,\mathbf{x}_j) }&\left[  \mathbf{g}(t_b+i\delta)\mathbf{d}_l^\ast e^{-t_x}e^{-i\chi_1(t_b,t_r)}\right.\nonumber\\ 
&\left.\,\times\,\mathbf{d}_j e^{-i\chi_2(t_b,t_r)}\right]
\end{align}
is governed by the prefactor $\mathbf{g}$, the ionization amplitude $\mathbf{d}_l^\ast e^{-t_x}$, the quantum mechanical phase factor $e^{-i\chi_1}$ acquired along the classical trajectory, and the recombination amplitude $\mathbf{d}_j e^{-i\chi_2}$.  For each possible birth site $\mathbf{x}_l$ and recombination site $\mathbf{x}_j$ in the lattice, the summation runs over all birth and recombination times $t_r,t_b$ that satisfy the saddle point conditions for a particular harmonic frequency $\omega$.

The propagator (\ref{WQCprop}) together with the saddle point equations (\ref{saddle}) and the interband current ({\ref{jeromw1}}) represent the WQC description of HHG in semiconductors. 
They reveal a complete and detailed picture of the physical mechanisms driving HHG in real and reciprocal space, summarized in figures \ref{fig1}(a) and (b), respectively. The empty circles 
in figure \ref{fig1}(a) represent the centers of the atomic unit cells $\mathbf{x}_l$, where $l = (l_x, l_y)$  in the 2D schematic. A Wannier basis function is located at each center. 
Initially, all Wannier sites of the valence band are filled. As all lattice sites are identical, it is sufficient to investigate $\mathbf{x}_l = 0$, see below (\ref{initial}). Following the
notation of our calculation we chose indices $l,j$ to represent birth and recombination sites, respectively. HHG proceeds in three steps. 

\textbf{Step 1 - creation of electron-hole pair by ionization.} At birth time $t_b$, a valence band electron localized at lattice site $\mathbf{x}_0$ transitions to the conduction band, and 
is localized at lattice site $\mathbf{x}_0+\mathbf{x}_l$. The tunneling probability is determined by the tunneling exponent $t_x$ and by the Wannier dipole moment $\mathbf{d}_l^*$, see figure 
\ref{fig1}(a). The potential energy experienced by the created electron-hole dipole in the laser field makes the effective ionization potential $E_g + F(t_b) x_l$ birth site dependent, see 
(\ref{stp}) and (\ref{tunnel}). In reciprocal space the electron transitions from valence to conduction band at the $\Gamma$-point at time $t_b$, see figure \ref{fig1}(b). Step 1 is of 
quantum mechanical nature. 

\textbf{Step 2 - electron-hole evolution in laser field.} The electron-hole pair quivers in the laser field. In real space it follows the classical trajectory $\boldsymbol{\xi}(t_b,t_r)$ 
in figure \ref{fig1}(a) until electron and hole revisit each other and are separated by $|\mathbf{x}_j|$ at time $t_r$, see (\ref{sk}). The propagation step is dominantly classical; of 
quantum mechanical nature are the phase $\chi_1(t_b,t_r)$ picked up between birth and recombination time, and the quasiclassical factor $\mathbf{g}$ coming from the quadratic expansion of the 
classical action $S$ about the classical trajectory. The shaded green area about the classical trajectory in figure \ref{fig1}(a) indicates the quantum correction up to second order. In  reciprocal space in figure \ref{fig1}(b) the electron-hole pair evolves from initial crystal momentum zero to saddle point crystal momentum $\mathbf{k}_s(t_b,t_r)$, defined below (\ref{saddle}). 

\textbf{Step 3 - recombination.} At time $t_r$ electron and hole recombine with probability amplitude $\mathbf{d}_j e^{-i \mathbf{k}_s(t_b,t_r)\cdot \mathbf{x}_j}$, see figure \ref{fig1}(a). The 
harmonic energy is given by the bandgap energy at $\mathbf{k}_s(t_b,t_r)$, see figure \ref{fig1}(b), plus the energy of the electron hole dipole in the field $F(t_r)$, see (\ref{st}). Due 
to the second term, harmonics with energies somewhat larger than the maximum bandgap can be generated. 

\section{Results}

\noindent
For the remainder of the paper, the WQC approach and its physical significance are explored within a 1D model system.  In this case the interband current, WQC propagator, and probability
amplitude reduce to scalars; namely $\tilde{j}_{er}$, $T_{jl}$, and $P_{jl}$. Specifically, we use a 1D delta function model potential, $V(x)=\Omega\sum_{n=-\infty}^\infty\delta[x-(n+1/2)a]$
with unit cell size $a$ and barrier penetration parameter $\Omega$. Details of the delta function model are given in appendix \ref{dltref}. For the investigated parameters the bandgap is well 
approximated by the nearest neighbour dispersion $\varepsilon(k)=E_g + \Delta[1-\cos(ka)]$, where $E_g$ is the minimum bandgap and $2\Delta$ represents the bandwidth.  We chose $a=7$ and 
considered two values $\Omega=0.5,1.5$ to model a weakly and tightly bound semiconductor, respectively.  The corresponding bandgap parameters are $E_g=0.141,0.269$; $\Delta=0.269,0.17$.  
Finally, for all runs we use a dephasing time $T_2 = T_0/2$ so that only returns within a single cycle are relevant. 

\begin{figure} [t]
\includegraphics[width=8.6cm]{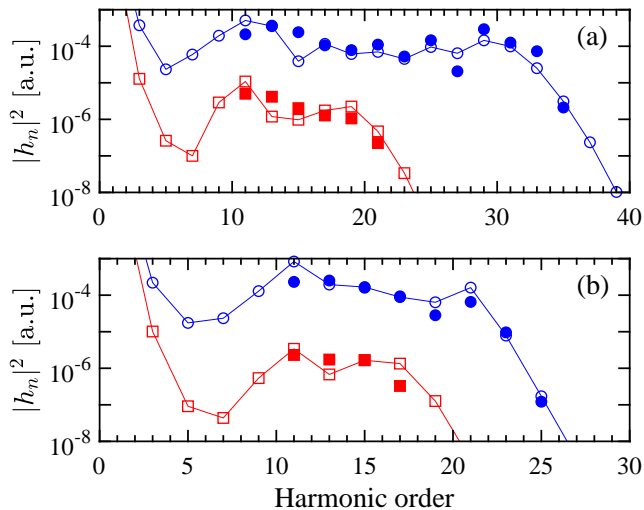} 
\caption{\label{fig2} Harmonic yield $|h_n|^2$ versus harmonic order $n$; $a = 7$, $T_2 = T_0/2$ (a,b); empty blue circles with lines (exact) and filled blue circles (WQC) refer 
to $\Omega = 0.5$, $\omega_0 = 0.01425$ ($\lambda = 3.2 \mu$m), and $F_0 = 0.0025$ in (a) and $F_0 = 0.0015$ in (b); empty red squares connected by lines (exact) and filled red squares (WQC) 
refer to $\Omega = 1.5$, $\omega_0 = 0.0285$ ($\lambda = 1.6 \mu$m), and $F_0 = 0.008$ in (a) and $F_0 = 0.005$ in (b); lines are used to guide the eye. }
% In (a) harmonic $n$=15 is highlighted;
% the discrepancy between exact and WQC result is larger for the weakly bound model ($\Omega=0.5$) than for the more tightly bound model ($\Omega=1.5$).}
\end{figure}

In figure {\ref{fig2}} the exact (quantum mechanical) harmonic spectrum, as obtained from numerical integration of (\ref{jerom}), is compared with the Wannier quasiclassical solution, 
(\ref{jeromw1}), (\ref{saddle}), (\ref{delta}), and (\ref{tprop}). For the exact approach we use $F(t) = F_0 \sin(\omega_0 t) \exp(-(t/\tau)^2)$ where $F_0$ is the maximum field 
strength, and the pulse duration, $\tau = 40 T_0$, is long enough to approach the continuous wave (cw) limit; $\omega_0$ is the laser center frequency and $T_0 = 2\pi / \omega_0$ denotes 
the optical cycle. We plot the harmonic intensity $\lvert h_n \rvert^2 = \int_{\omega_{-}}^{\omega_{+}} 
d\omega \lvert \tilde{{j}}_{er}(\omega) \rvert^2$ integrated over the frequency interval $\omega_{\pm} = (n\pm 1/2) \omega_0$.

\begin{figure}[t!]
\begin{center}
\includegraphics[width=8.6cm]{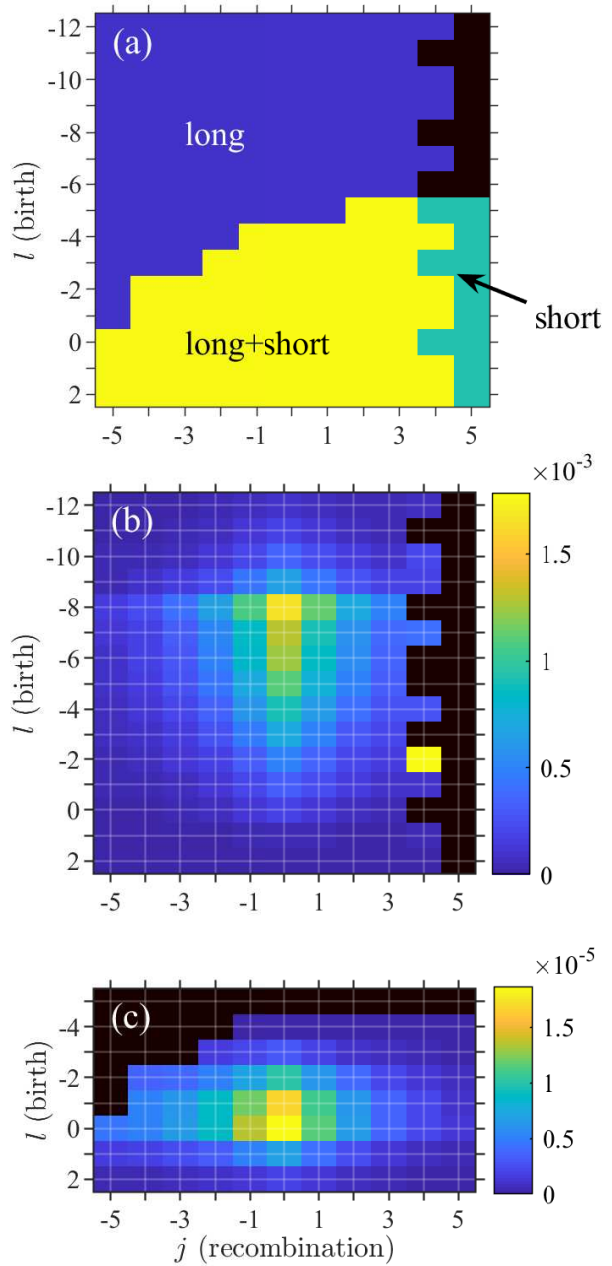} 
\caption{\label{fig3} 
The contribution of the long and short classical trajectories to the probability amplitude $|P_{jl}|$ for harmonic order $n=15$ in a wide-band semiconductor; parameters $a=7$, $\Omega=0.5$, $\omega=0.01425$, and $F_0=0.0025$ corresponding to filled blue circles in figure \ref{fig2}(a).  Figure (a) depicts the combinations of birth ($l$) and recombination ($j$) site indices for which each trajectory exists and contributes to $P_{jl}$; black regions indicate no solution.  Figure (b) shows the contribution of the long trajectory to $|P_{jl}|$, while figure (c) shows the contribution from the short trajectory. Note that the values of the colorscale differ by two orders of magnitude in (b) and (c).}
\end{center}
\end{figure}

\begin{figure} [!t]
\includegraphics[width=8.7cm]{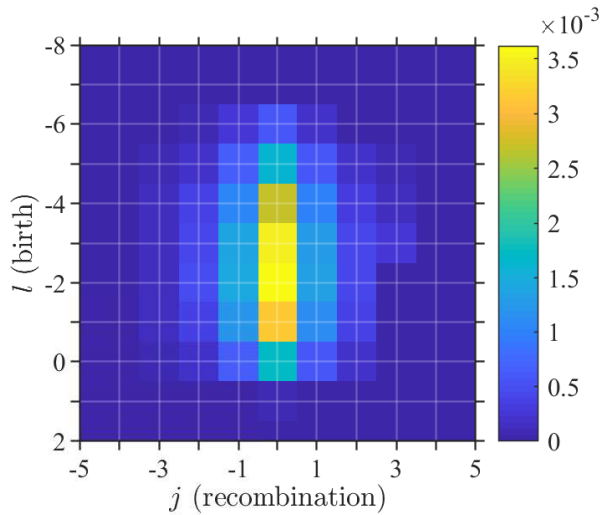} 
\caption{\label{fig4} Probability amplitude $|P_{jl}|$ versus birth ($l$) and recombination ($j$) site indices for harmonic order $n=15$ in a narrow-band semiconductor; parameters $a=7$, $\Omega=1.5$, $\omega=0.0285$, and $F_0=0.008$ corresponding to filled red circles in figure \ref{fig2}(a). Here we plot the total probability amplitude $|P_{jl}|$, but note that the long trajectory is dominant; the individual contributions are similar to the behaviour depicted for the wide-band semiconductor in figure \ref{fig3}(b).}
\end{figure}

For the WQC calculation we assume the continuous wave limit, $F(t) = F_0 \sin(\omega_0 t)$, in order to facilitate interpretation of the results. Equation (\ref{tprop}) has been derived 
for finite pulses employing the Fourier transform. For a transition to the cw limit, the Fourier transform has to be replaced by a Fourier series; as a result, $\omega \rightarrow n \omega_0$, 
pre-factor $g \rightarrow g / (2 \pi T_0)$, where the $ 1/(2\pi)$ comes from the 1D nature of our model. The harmonic yield becomes $\lvert h_n \rvert^2 = \lvert 
\tilde{{j}}_{er}(n \omega_0) \rvert^2$ with $T_{jl}$ given by the WQC propagator (\ref{tprop}).

In figure \ref{fig2} the blue empty circles (exact) and blue filled circles (WQC) refer to results for the weakly bound model semiconductor, with $\Omega = 0.5$, $\omega_0 = 0.01425$. Red empty squares (exact) and red filled squares 
(WQC) refer to the tightly bound semiconductor, with $\Omega = 1.5$, $\omega_0 = 0.0285$. Plots with the same symbols in figures \ref{fig2}(a) and (b) correspond to the same values of $\Omega$ and $\omega_0$, but differ in $F_0$.

The WQC approach agrees well with the exact solution, with most data points being off by less than a factor 2. Even the first 1-2 cutoff harmonics are described fairly well, which 
demonstrates that they are of quasi-classical origin. The good agreement allows us to interpret semiconductor quantum dynamics such as ionization, electron/hole transport, and HHG in terms 
of classical trajectories. The quantum contributions to HHG are captured by the tunneling exponent $t_x$, by the pre-exponential factor $g$ in (\ref{tprop}), and by the Wannier dipole
moments in (\ref{jeromw}). 

A few points disagree by a larger factor of up to 6.  In particular, figure \ref{fig2}(a) shows that the WQC result for harmonic $n=15$ exhibits larger discrepancy for the weakly bound
semiconductor ($\Omega=0.5$) compared to the more tightly bound semiconductor ($\Omega=1.5$). The reason for this behaviour is identified in figure \ref{fig3} and will be discussed later. 

Numerical solution of the full saddle point equations reveals two distinct classical trajectories that contribute to the probability amplitude $P_{jl}$; one long trajectory and one short. 
Moreover, each solution exists for only certain combinations of birth ($l$) and recombination ($j$) lattice sites.  Figure \ref{fig3} shows the contributions arising from the different 
classical trajectories for the fifteenth harmonic $(n=15)$ with $\Omega = 0.5$, $F_0 = 0.0025$, corresponding to the filled blue circles in figure \ref{fig2}(a).  Figure  \ref{fig3}(a) 
depicts the regions in the $j$-$l$ plane where each trajectory contributes to $|P_{jl}(n=15)|$.  No solution exists for the dark region in the top-right, and the probability amplitude here 
is zero.  Figures \ref{fig3}(b) and (c) show the individual contributions to the probability amplitude from the long and short trajectories, respectively.  The long trajectory is dominant, 
as the electron-hole pair is born close to the field peak, whereas the short trajectory is born closer to the nodal point. This outweighs the effect of the short dephasing time, which favors 
the short trajectory. As a result, the contribution of each data point to the WQC propagator is dominantly determined by a factor $\sim ge^{-t_x}$ of a single (long) trajectory. The full
probability amplitude $|P_{jl}(n=15)|$ is essentially identical to figure \ref{fig3}(b).

In figure \ref{fig4} the total probability amplitude for the fifteenth harmonic $\lvert P_{jl}(n=15) \rvert$ is plotted as a function of birth and recombination site indices $l,j$ for 
$\Omega = 1.5$, $F_0 = 0.008$, which corresponds to the filled red squares in figure \ref{fig2}(a). For this system the long trajectory is also dominant, and analysis of the individual 
contributions would reveal a picture qualitatively similar to figure \ref{fig3}.

In both figures \ref{fig3} and \ref{fig4}, harmonic $n=15$ has been selected, as the WQC result for the weakly bound semiconductor exhibits a more pronounced difference, while it agrees well 
for the tightly bound semiconductor. For both systems, the maximum probability is shifted towards negative birth site indices; it is more likely for electron and hole to be born apart than 
at the same site. Tunnel ionization probability is determined by $e^{-t_x}$ and by birth dipole moment $d_l^*$. The tunnel exponent $t_x$ depends on the ionization potential $E_g + F(t_b) 
x_l$, see (\ref{tprop}). Thus, for positive field the electron-hole pair gains energy when born at increasingly negative distances which reduces $t_x$. When $-x_l = E_g/F(t_b)$, $t_x$ vanishes;
in other words, the valence and conduction band levels separated by $-l$ sites align, and the electron hops from the valence to the conduction band site. The penalty to be paid is a rapidly
dropping dipole moment $d_l$. As such, the birth site index at which ionization is maximum is determined by a tradeoff between tunnel exponent and Wannier dipole moment. The dipole elements 
for the parameters of figure \ref{fig3}(a) drop more slowly with increasing $\lvert l \rvert$ than for (b); see appendix \ref{dltref}. Therefore, the site of highest ionization probability 
is shifted more strongly towards negative $l$. Recombination is most probable for $j=0$ in figures \ref{fig3}(a) and (b) which is consistent with previous findings \cite{Osika2017}. The drop 
in probability for increasing $j$ is due to $d_j$, which is why $\lvert P_{jl} \rvert$ extends to larger $j$ in figure \ref{fig3}(b). 

The results in figures \ref{fig3} and \ref{fig4} are displayed for birth times in the positive field cycle $0 \le t_b \le T_0/2$; the negative half cycle would show the same picture, but
mirrored about the $x-$ and $y-$axis ($j,l \rightarrow -j,-l$). 

Recall that exact and quasiclassical results do not agree well for harmonic $n=15$ in figure \ref{fig2}(a) ($\Omega = 0.5$). The reason is found in figure \ref{fig3}(b); disagreement 
is due to the point $(j,l) = (4,-2)$ that exhibits unusually high probability. We find that at this point $k_s$ is approximately zero, and therewith $|\mathcal{H}|\approx 0$. 
Since $g\propto 1/\sqrt{|\mathcal{H}|}$, this leads to a large value of the prefactor $g$. This behaviour indicates that the quadratic saddle point expansion is no longer sufficient and 
the next higher order term(s) must be included. In contrast, agreement for harmonic $n=15$ in figure \ref{fig2}(a) for $\Omega = 1.5$ is good. This is consistent with the fact that in figure 
\ref{fig4}, $k_s \approx 0$ does not occur in areas of high probability. 

Finally, the WQC method hinges on saddle point integration which works well when the exponent is rapidly oscillating. This is fulfilled for wide-band semiconductors with large bandwidth 
($\Delta$) and in the long wavelength limit. When transitioning to smaller $\Delta$ (dielectrics) and shorter wavelengths, saddle point integration is expected to fail at some point. This will
be subject to further research. Also, it is generally possible for transitions involving higher conduction bands to contribute to the harmonic spectrum, but this is beyond the scope of the two
band model considered here.

\vspace{1em}
\section{Conclusion}
\noindent

In summary, we have shown that the full quantum dynamics driving HHG in wide band materials, such as semiconductors, can be quantitatively explained in terms of quasi-classical trajectory
propagation. The physical insight offered by trajectory analysis will prove useful for optimization and design of strong field and attosecond experiments and for the development of novel
diagnostic applications of HHG, such as reconstruction of the dipole moment \cite{Zhao2019}. We believe that our approach presents a versatile tool for investigating open issues in strong 
field solid state physics, such as the role of noise and many-body effects in strong field processes. Beyond that, quantitatively accurate quasi-classical analysis should be of interest 
for a wider range of topics in material science. 

P.\! B.\! Corkum acknowledges the support of AROSR grant number FA9550-16-0109. G. Ernotte was supported by the Vanier Canada Graduate Scholarship program.
\vspace{0.3cm}
%\newpage
%\clearpage

\appendix

\section{Hessian}
\label{app_Hess}

\noindent
Here we provide expressions for the determinant of the Hessian $\mathcal{H}_{ij}=\partial^2\varphi/\partial_i\partial_j$ appearing in (\ref{WQCprop}). Evaluation of the second derivatives 
yields 
\begin{widetext}
\begin{align}
\lvert \mathcal{H} \rvert = 
\left \lvert {\begin{array}{ccccc}
 \mathbf{F}(t') \cdot\mathbf{v}(\mathbf{k}(t',t)) & - \mathbf{F}(t)\cdot \mathbf{v}(\mathbf{k}(t',t)) &  v_x(\mathbf{k}(t',t)) &  v_y(\mathbf{k}(t',t)) &  v_z(\mathbf{k}(t',t)) \\
+ \dot{\mathbf{F}}(t') \cdot\mathbf{x}_l & & & & \\[5pt]
- \mathbf{F}(t)\cdot \mathbf{v}(\mathbf{k}(t',t)) & \mathbf{F}(t)\cdot \mathbf{v}(\mathbf{k}) - \dot{\mathbf{F}}(t)\cdot \mathbf{x}_j - & -v_x(\mathbf{k}) +  & -v_y(\mathbf{k}) +  & -v_z(\mathbf{k}) +  \\
~& F_i(t) D_{ij}(t',t) F_j(t) & F_i(t) D_{ix}(t',t) & F_i(t) D_{iy}(t',t) & F_i(t) D_{iz}(t',t) \\[5pt]
 v_x(\mathbf{k}(t',t)) & -v_x(\mathbf{k}) +  F_i(t) D_{xi}(t',t) & -D_{xx}(t',t) & -D_{xy}(t',t) & -D_{xz}(t',t) \\[5pt]
 v_y(\mathbf{k}(t',t)) & -v_y(\mathbf{k}) + F_i(t) D_{yi}(t',t) & -D_{yx}(t',t) & -D_{yy}(t',t) & -D_{yz}(t',t) \\[5pt]
 v_z(\mathbf{k}(t',t)) & -v_z(\mathbf{k}) + F_i(t) D_{zi}(t',t) & -D_{zx}(t',t) & -D_{zy}(t',t) & -D_{zz}(t',t) \\[5pt]
\end{array} } \right \rvert_{(t'=t_b+i\delta, t=t_r, \mathbf{k} = \mathbf{k}_s)} \text{,}
\label{hessian1}
\end{align}
\end{widetext}
Using linear dependence between column 2 and columns 3,4, and 5, see the supplement of \cite{Uzan2019}, the 
determinant can be simplified to 
%\clearpage
\begin{widetext}
\begin{align}
\lvert \mathcal{H} \rvert = 
\left \lvert {\begin{array}{ccccc}
\mathbf{F}(t')\cdot \mathbf{v}(\mathbf{k}(t',t))+\dot{\mathbf{F}}(t') \cdot\mathbf{x}_l  & 0 &  v_x(\mathbf{k}(t',t)) &  v_y(\mathbf{k}(t',t)) &  v_z(\mathbf{k}(t',t)) \\[5pt]
- \mathbf{F}(t)\cdot \mathbf{v}(\mathbf{k}(t',t)) & -\dot{\mathbf{F}}(t)\cdot \mathbf{x}_j & -v_x(\mathbf{k})\ +  & -v_y(\mathbf{k})\ +  & -v_z(\mathbf{k})\ +  \\
 & & F_i(t) D_{ix}(t',t) & F_i(t) D_{iy}(t',t) & F_i(t) D_{iz}(t',t) \\[5pt]
 v_x(\mathbf{k}(t',t)) & -v_x(\mathbf{k}) & -D_{xx}(t',t) & -D_{xy}(t',t) & -D_{xz}(t',t) \\[5pt]
 v_y(\mathbf{k}(t',t)) & -v_y(\mathbf{k}) & -D_{yx}(t',t) & -D_{yy}(t',t) & -D_{yz}(t',t) \\[5pt]
 v_z(\mathbf{k}(t',t)) & -v_z(\mathbf{k}) & -D_{zx}(t',t) & -D_{zy}(t',t) & -D_{zz}(t',t) \\[5pt]
\end{array} } \right \rvert_{(t'=t_b+i\delta, t=t_r, \mathbf{k} = \mathbf{k}_s)} \text{.}
\label{hessian2}
\end{align}
\end{widetext}
Here, $i,j\!\!\in\!\!\{x,y,z\}$, summation is implied when indices $i$ or $j$ are repeated, $D_{ij} = \int_{t'}^t d\tau \beta_{ij}(\mathbf{k}(t'',t))$, $\beta_{ij} = \partial_{k_i} 
v_j(\mathbf{k})$, and  $\dot{\mathbf{F}}(t)=\partial_t\mathbf{F}(t)$. 
For completeness $\lvert \mathcal{H} \rvert$ is given for a general field $\mathbf{F}(t)$; for the case treated here, set $F_y = F_z = 0$. 
To leading order $\lvert \mathcal{H} \rvert = v_x(\mathbf{k}) \mathbf{f}(t',t,\mathbf{k}) + \dot{\mathbf{F}}(t) \cdot\mathbf{x}_l h(t',t,\mathbf{k})$, where $h, \mathbf{f}$ are minors of 
$\lvert \mathcal{H} \rvert$. For completeness, we have included time derivatives of the laser field which are however small in the long wavelength limit. As a result the leading order 
term is $\lvert \mathcal{H} \rvert = v_x(\mathbf{k}) \mathbf{f}(t',t,\mathbf{k})$. 

\section{Delta function potential}
\label{dltref}

\begin{figure}[t!]
\includegraphics[width=8.6cm]{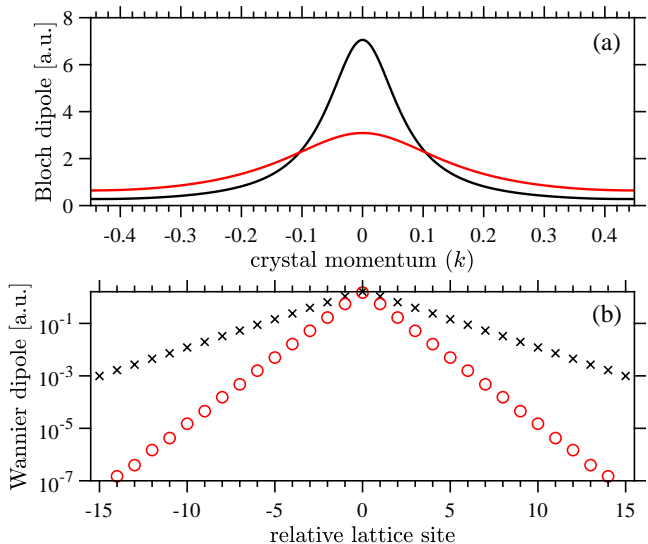} 
\caption{\label{figB} (a) Bloch dipole transition elements Im$[d^\ast(k)]$ versus $k$; (b) Wanner dipole transition elements $d_j$ versus $j$ which represents the difference in lattice sites at
which electron and hole are born. 1D model parameters: $a=7$, $\Omega = 0.5$ (black), $\Omega = 1.5$ (red).}
\end{figure}

\noindent
The WQC approach and its physical significance are explored by means of a 1D delta-function model potential, $V(x) = \Omega 
\sum_{n=-\infty}^{\infty} \delta[x-(n+1/2)a]$ with unit cell size $a$ and barrier penetration parameter $\Omega$. For the investigated parameters the bandgap is well approximated by 
the nearest neighbor approximation, $\varepsilon = E_g + \Delta [1 - \cos(ka) ]$, where $E_g$ is the minimum bandgap and $2 \Delta$ represents the bandwidth. 

The binding energy is determined by $2 E_m = K_m^2$, where $m=v,c$ and $K_m$ is determined by 
\begin{align} 
\cos(ka) = \cos(K_m a) + \frac{\Omega}{K_m} \sin(K_m a) \text{.}
\label{senergy}
\end{align} 
The wavefunction is given by 
\begin{align}
\Phi_{{m},{k}}({x}) & = \sqrt{\frac{1}{a}} u_{m,k}(x) \exp(ikx) \label{swf} \\ 
u_{m,k}(x) & = A_m(k) \left[ e^{i(K_m-k)x} + r_m e^{-i(K_m+k)x} \right]   \nonumber \\
A_m(k) & = 1 / \sqrt{1 + r_m^2 + 2 r_m \sin(K_m a) / (K_m a)}  \nonumber \\
r_m(k) & = \frac{\sin[(K_m-k)a/2]}{\sin[(K_m + k)a/2]} \nonumber
\end{align}
From the wavefunction the Bloch dipole moment is found to be 
\begin{align}
d_{cv}(k) & = d^*(k) = i \frac{2 A_c A_v}{E_c - E_v} \times \label{sdipole} \\
& \left\{ [ (K_v-k)r_c - (K_v+k)r_v ] \frac{\sin[ (K_v+K_c) a/2]}{(K_v+K_c) a} +  \right. \nonumber \\
& \left. + [ (K_v-k) - (K_v+k)r_v r_c ] \frac{\sin[ (K_v-K_c) a/2]}{(K_v-K_c) a} \right\} \nonumber \text{.} 
\end{align}

We chose $a\!\!=\!\!7$ and $\Omega\!\!=\!\!0.5,1.5$ to model a weakly and red more tightly bound semiconductor, respectively. The corresponding bandgap parameters are 
$E_g\! =\! 0.141 , 0.269$; $\Delta \! =\! 0.269 , 0.17$. The Bloch dipole elements $d(k)$ and Wannier dipole elements $d_j$ are plotted in figure \ref{figB}. As expected, $d_j$ drops 
faster for the more tightly bound model. Finally, we have chosen the coordinate center at the point of inversion symmetry which corresponds with choosing a maximally 
localized Wannier basis \cite{Kohn1959}. For this choice the diagonal (intraband) dipole moments are zero and the phase of the interband dipole moment is constant.

\bibliography{manuscript_jmp}

\end{document}